\documentstyle[12pt,epsfig]{article}
\title{Brane world cosmological constant in the models with large 
extra dimensions.}
\author{ Andrey Neronov\\
Theoretische Physik, \\
Universit\"at M\"unchen, \\
Theresienstr. 37/III, \\
80333, Munich, Germany}

\begin{document}

\maketitle

\begin{abstract}
We consider ``brane universe'' with nonzero tension 
in the models with large extra dimensions. We find exact solutions of 
higher-dimensional Einstein equations with single 
flat Minkowsky brane of arbitrary 
large tension (or brane cosmological constant) 
and compact extra dimensions. The brane curves the bulk space-time in a small
region around it. There is no fine tuning of 
energy scales in our model. 
\end{abstract}

Phenomenological models with large  
extra space-time dimensions \cite{dimopoulos} have attracted
considerable attention recently.
Like in conventional Kaluza-Klein models, in the
models with large extra dimensions 
higher-dimensional space-time is supposed to be a direct product 
${\cal M}^{4+n}=R^4\times K^n$ of a four-dimensional plane with a compact 
manifold $K^n$ (for example, an $n$-dimensional sphere $S^n$ of radius 
$R$). The metric 
on ${\cal M}^{4+n}$ is just a sum of flat Minkowsky metric on 
$R^4=\{x^\mu\}, \mu=0,1,2,3$ and a standard metric on 
$K^n=\{y^a\}, a=1,..,n$
\begin{equation}
ds^2=\eta_{\mu\nu}dx^\mu dx^\nu+g_{ab}(y)dy^a dy^b
\end{equation} 
The key difference of models with large, or infinite extra dimensions 
from the conventional Kaluza-Klein models is that matter fields of the 
Standard 
Model of particle physics 
are confined to a four-dimensional surface (3-brane) $M^4$ embedded into 
${\cal M}^{4+n}$. The brane $M^4$ can be viewed  as a 
topological defect in higher-dimensional space. The Standard 
Model particles can be localized on this defect either due to some 
field-theoretical 
mechanism \cite{rubakov,akama} or in a way supposed by 
String Theory \cite{string}.

If matter fields of the Standard Model are localized on the brane, the size $R$
of extra dimensions is not limited to be very small. If $R$ 
is large enough, the higher-dimensional Planck mass 
$M_{pl(4+n)}$ which is related to the four-dimensional one 
$M_{pl(4)}\sim 10^{16}$ TeV as
\begin{equation}
M_{pl(4+n)}^{n+2}=M_{pl(4)}^2R^{-n}
\end{equation}
can be as small as several TeV, which opens up a possibility of testing the 
presence of extra dimensions and quantum gravity effects in future collider 
experiments. 

Even if the higher-dimensional Planck mass is much larger than TeV, 
phenomenological models with brane universe have generically another
testable feature \cite{sundrum,dobado}.
The presence of the brane brakes spontaneously a part of translational 
invariance in higher-dimensional space-time. Goldstone bosons associated
to this symmetry breaking behave themselves in first approximation 
as massless scalar particles localized on the brane. These particles interact 
with Standard Model fields. Such interaction produces an essential 
contribution to particle collisions on energy scales higher or approximately 
equal to $\mu=\Lambda^{1/4}$ where $\Lambda$ is the typical energy density 
 on the brane. Even if $M_{pl(4+n)}$ is essentially 
higher than 1 TeV, the energy scale $\mu$ can be accessible to  accelerators.
The lower bound on $\mu$ 
imposed by requirement of consistency with electroweak
physics \cite{creminelli} and astrophysical bounds 
\cite{kugo} turns out to be 
\begin{equation}
\label{lower}
\mu\geq 10^{-1} \mbox{ TeV}
\end{equation}  
The scale of the energy density on the brane $\Lambda$ must, therefore, be 
much larger than the natural cosmological energy density scale
\begin{equation}
\Lambda=\mu^4\gg \rho_{cr}=\frac{3H^2}{8\pi G_4}\sim 10^{-58} \mbox{ TeV}^4
\end{equation}  
where $\rho_{cr}$ is the critical energy density of the universe, $H$ is
the Hubble constant and $G_4$ is the four-dimensional gravitational constant. 
All the energy density $\rho\sim \rho_{cr}$ 
provided by the visible matter in the universe is 
 just a small perturbation of the energy density $\Lambda$ of the brane. 
The energy density $\Lambda$ can be ascribed, for example to the  
topological defect  itself or to the vacuum energy of the 
Standard Model fields localized on the brane. 

The higher dimensional space-time ${\cal M}^{4+n}$ 
must be a solution of higher-dimensional 
Einstein equations with compact extra dimensions and a brane with flat 
Minkowsky geometry and nonzero tension $\Lambda$. The general maximally 
symmetric stress-energy tensor of the brane has the form
\begin{equation}
\label{tmn}
T_{\mu\nu}\sim -\Lambda \gamma_{\mu\nu}\delta(M^4)
\end{equation}
where $\gamma_{\mu\nu}$ is the induced metric on the brane 
which is equal to $\eta_{\mu\nu}=\mbox{diag}(-1,1,1,1)$ for the Minkowsky
brane.

Since the 
stress-energy tensor of the brane is not zero, the brane must curve  bulk 
space-time around it. In general, it is not 
obvious that there exists a {\it static} solution of higher-dimensional 
Einstein equations with {\it single} Minkowsky brane of nonzero tension 
$\Lambda$ 
and {\it compact} extra dimensions. 
Moreover, we know that in the 
absence  of extra dimensions the presence of nonzero stress-energy
tensor of the form (\ref{tmn})  leads to exponential expansion of the
universe. In the presence of extra dimensions large vacuum energy can 
curve  extra dimensions, leaving the four ``physical''
dimensions flat \cite{rubakov2}. There 
exists a number of 
static  solutions of higher-dimensional 
Einstein equations with bulk cosmological 
constant and a brane of nonzero tension 
\cite{neronov}-\cite{hidden}. But 
in the known static
solutions the extra dimensions are either infinite 
\cite{rs2,gregory,shaposhnikov}
or they are finite but 
there is another ``hidden'' brane \cite{rs1,hidden} or 
curvature singularity of another kind \cite{singularity,sing}
at a finite distance from the ``observable'' four-dimensional brane.  
It is possible to get rid of the ``hidden'' brane making the compact
extra dimensions nonstatic \cite{neronov}. 

In what follows we present solutions of higher-dimensional Einstein equations
with a flat brane of nonzero tension $\Lambda$ and compact extra dimensions.
The scale of $\Lambda$ is not fine tuned neither to the higher-dimensional
Planck scale $M_{pl(4+n)}^4$ nor to the Kaluza-Klein scale $R^{-4}$ and is 
defined by internal properties of the brane.

Typical size of the space-time region  
around the brane in which the background geometry  is 
essentially deformed by the presence of the brane is much smaller than
both  the higher-dimensional Planck length and the size of the extra 
dimensions even if the brane tension is as high as $M_{pl(4+n)}^4$. 
Therefore, in description of classical and quantum fluctuations of the brane  
around its equilibrium position we can use the approximation of test brane 
on a fixed background space-time in a way supposed in 
\cite{sundrum,dobado,creminelli,kugo}.

We consider a space-time ${\cal M}^{4+n}=R^4\times S^n$  
with compact extra dimensions with topology
of the sphere $S^n$. Let us restrict our attention to the case $n=4$, 
although the construction is easily generalized for arbitrary $n\ge 4$. 
The set of coordinates on $M^4\times S^4$ is 
\begin{equation}
x^A=(\underbrace{x^\mu,}_{R^4}\underbrace{r, \psi, \theta, 
\phi}_{S^4})
\end{equation}
In the absence of the brane the metric on $S^4$ is the standard metric on 
a sphere of radius $R$
\begin{equation}
\label{s4}
ds^2=\eta_{\mu\nu}dx^\mu dx^\nu +\left(1-\frac{r^2}{R^2}\right)d\psi^2+
\left(1-\frac{r^2}{R^2}\right)^{-1}dr^2+r^2(d\theta^2+\sin^2\theta d\phi^2)
\end{equation}
The coordinate $r$ is defined within the interval $r\in (0, R)$, 
the coordinate 
$\psi$ is periodic with the period $2\pi R$ and $(\theta, \phi)$ are the 
standard angular coordinates on a two-dimensional sphere. The coordinate 
system $(r,\psi, \theta, \phi)$ becomes singular at $r=0, R$. In the vicinity 
of $r=0$ the radius of two-dimensional sphere parameterized by 
$(\theta, \phi)$ goes to zero so that the region $0\le r< \hat r$ for 
some small $\hat r$ has topology $D^3\times S^1$ of direct product 
of three-dimensional disk parameterized by $(r, \theta, \phi)$
and a circle parameterized by $\psi$. Singularity of coordinate 
system at $r=0$ is just the singularity of three-dimensional spherical
coordinates at $r=0$. In the vicinity of $r=R$ radius of the 
circle $S^1$ parameterized by $\psi$ goes to zero and topology 
of a region $\hat r<r\le R$ is $S^2\times D^2$ where $S^2$ is parameterized by
$(\theta, \phi)$ and two-dimensional disk $D^2$ is parameterized by 
$(r,\psi)$. Singularity of the coordinate system at $r=R$ is the 
same as at the origin of polar coordinates on a two-dimensional plane.  
The whole sphere $S^4$ is a connected sum
$S^4=(D^3\times S^1)\# (S^2\times D^2)$ of the above two manifolds 
with boundaries.  

The space-time (\ref{s4}) is a solution of the eight-dimensional 
Einstein equations with stress-energy tensor
\begin{eqnarray}
\label{str}
T^\mu_\nu&=&-\rho\delta^\mu_\nu\nonumber\\ 
T^a_b&=&-w\rho\delta^a_b
\end{eqnarray}
where the energy density $\rho$ and parameter $w$ which describes the
equation of state are
\begin{eqnarray}
\label{w}
\rho&=&\frac{3}{4\pi G_8 R^2}\nonumber \\ 
w&=&1/2
\end{eqnarray}
Here $G_8$ is the eight-dimensional gravitational constant.

Let us now consider a brane $M^4$ embedded 
into ${\cal M}^8$ as $M^4:\{r<\hat r\}$ where $\hat r\ll R$ is a small brane 
thickness.  Presence of the brane destroys the maximally symmetric 
form of the bulk metric (\ref{s4}). A general form of metric in the bulk 
is now 
\begin{equation}
\label{general}
ds^2=e^{A(r)}\eta_{\mu\nu}dx^\mu dx^\nu+e^{B(r)}\psi^2+e^{C(r)}dr^2+
r^2(d\theta^2+\sin^2\theta d\psi^2)
\end{equation}
with arbitrary functions $A(r), B(r), C(r)$. Outside the brane the metric
must be a solution of eight-dimensional Einstein equations with the same 
bulk stress-energy tensor (\ref{str}). The most simple choice is 
\begin{eqnarray}
\label{r>r}
ds^2_{r>\hat r}=\eta_{\mu\nu} dx^\mu dx^\nu+
\left(1-\frac{r_0}{r}-\frac{r^2}{R^2}\right)d\psi^2+
\left(1-\frac{r_0}{r}-\frac{r^2}{R^2}\right)^{-1}dr^2+\nonumber\\
r^2(d\theta^2+\sin^2\theta d\phi^2)
\end{eqnarray}  
the sum of Minkowsky metric along $x^\mu$ directions and 
Eucledian Schwarzschild-deSitter black hole with gravitational radius $r_0$. 
The coordinate 
$\psi$ is defined now with the period
$2\pi \tilde R$ where $\tilde R$ is chosen in such a way that the 
space-time has no conical singularity at $r=r_1$ where $r_1$ is 
the larger root of 
cubic equation $r^3+R^2(r_0-r)=0$. If we are interested in the case 
when both the thickness  $\hat r$ and the gravitational radius 
$r_0$ are much smaller than $ R$, then  $r_1\approx \tilde R\approx R$.

The metric ``inside'' the brane in the region
$r<\hat r$ depends on  details of the brane structure and behavior
of the functions $A(r), B(r), C(r)$ (\ref{general}) in the vicinity of $r=0$
can be different for different types of four-dimensional topological defects
in higher-dimensional space. The metric in the region $0<r<\hat r$ inside the 
brane must satisfy the boundary condition
\begin{equation}
\label{r=r}
ds^2_{r=\hat r}=\eta_{\mu\nu}dx^\mu dx^\nu+
\left(1-\frac{r_0}{\hat r}-\frac{\hat r^2}{R^2}\right)d\psi^2+
\hat r^2(d\theta^2+\sin^2\theta d\phi^2)
\end{equation}
in order to match the outer metric (\ref{r>r}) continuously.
For example, let us take the metric  
\begin{equation}
\label{r<r}
ds^2_{r<\hat r}=\eta_{\mu\nu}dx^\mu dx^\nu+
\left(1-\frac{r^2}{a}\right)d\tilde \psi^2+
\left(1-\frac{r^2}{a}\right)^{-1}
dr^2+r^2(d\theta^2+\sin^2\theta d\phi^2)
\end{equation}
In this case the complete space-time with the brane is constructed as follows.
We cut a small region $0<r<\hat r$ of the four-dimensional sphere $S^4$ 
of large radius $R$ and glue a space  locally  isometric to a smaller 
sphere of radius $\sqrt{a}$ on inside $r=\hat r$ as it is 
schematically presented on Fig. 1. 

\begin{figure}
\begin{center}
\epsfig{file=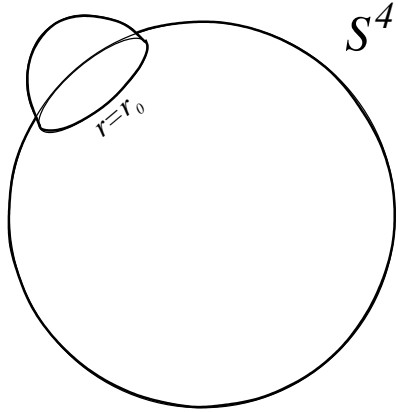}
\end{center}
\caption{The geometry of the sphere $S^4$ deformed by the presence of the 
brane inside $r=\hat r$.}
\label{fig:1}
\end{figure}

The coordinate $\tilde \psi$ is defined with an arbitrary period 
$\tilde \psi\in (0, 2\pi L)$ and the relation between $a$ and $L$ is found
from the requirement that the induced metric on the surface $r=\hat r$ is 
equal to (\ref{r=r})   
\begin{equation}
L^2\left(1-\frac{\hat r^2}{a}\right)=
R^2\left(1-\frac{r_0}{{\hat r}}-\frac{\hat r^2}{R^2}\right)
\end{equation}
 The matter stress-energy tensor in the region $0<r<\hat r$ can be calculated 
from the Einstein  equations
\begin{equation}
T^A_B=\frac{1}{8\pi G_8}\left(R^A_B-\frac{1}{2}\delta^A_B R\right)
\end{equation}
It has the form (\ref{str}) with 
\begin{eqnarray}
\label{bul}
\rho_{r<\hat r}&=&\frac{3}{4\pi G_8 a}\nonumber\\
w_{r<\hat r}&=&1/2
\end{eqnarray} 
The surface stress-energy tensor at $r=\hat r$
can be found from Israel junction conditions across $r=\hat r$ 
which relate it to the jump of extrinsic curvature $K^A_B$ 
of the surface $r=\hat r$
\begin{equation}
\label{israel}
T^A_B=\frac{1}{8\pi G_8}\left( [K^A_B]-\delta^A_B[K]\right)
\end{equation}
(square brackets denote the jump $[K]=K(\hat r+0)-K(\hat r-0)$).
Substituting the metrics (\ref{r<r}) and (\ref{r>r}) into (\ref{israel})
we find 
\begin{equation}
\label{sur}
T^\mu_{\nu(r=\hat r)}=\frac{3r_0R^2(2R-L)+
2\hat rR(L^2+2RL-3R^2)+
6\hat r^3(R-L)}{16\pi G_8\hat r^2L^2 R
\sqrt{1-\hat r^2/a}}\delta^\mu_\nu
\end{equation}
The effective four-dimensional stress-energy tensor of brane  
is obtained by integrating the eight-dimensional stress-energy tensor over 
the brane volume 
\begin{equation}
\label{4d}
{\cal T}^\mu_\nu=4\pi \int\limits_{0}^{\hat r} r^2dr\int\limits_0^{2\pi L} 
d\psi 
T^\mu_{\nu(r<\hat r)}+
4\pi\hat r^2\int\limits_0^{2\pi L}d\psi \sqrt{1-\hat r^2/a} 
T^\mu_{\nu(r=\hat r)}
\end{equation}
Substituting (\ref{bul}) and (\ref{sur}) into (\ref{4d}) we get
\begin{equation}
{\cal T}^\mu_\nu=-\Lambda\delta^\mu_\nu
\end{equation}
where  the brane tension, or the cosmological constant on the brane is 
\begin{equation}
\label{lam}
\Lambda=\frac{\pi}{G_8}\left(
\left(\frac{3}{R}-\frac{1}{L}\right)\hat r^3+
\left(L-2R+\frac{R^2}{L}\right)\hat r+
\left(\frac{3}{2}R-L\right)r_0
\right)
\end{equation}
Depending on the values of parameters $L, \hat r, r_0$ the brane
tension can be both positive or negative. We consider several particular
cases. If take $L\rightarrow 0$ so that the size 
of $S^1$ parameterized by $\psi$ is very small at $r=0$,
the brane tension is approximately
\begin{equation}
\label{one}
\Lambda\approx\frac{\pi\hat r R^2}{G_8 L}
\end{equation}  
If we choose $L\approx R$ so that the size of $S^1$ is large at $r=0$, but
becomes small at $r=\hat r$ if $\hat r\approx r_0$, the brane tension is
\begin{equation}
\label{two}
\Lambda\approx\frac{\pi r_0 R}{2G_8}
\end{equation} 
Thus, the cosmological constant on the brane depends on the internal
structure of the brane. 

Let us discuss the question of fine tuning of parameters in our model. 
Within four-dimensional General Relativity cosmological constant 
must be fine tuned to zero in order to explain flat Minkowsky geometry 
of space-time. In the models with extra dimensions the presence
of nonzero cosmological constant on the brane (\ref{tmn}) 
does not necessary cause the exponential expansion of the brane universe.
 The brane tension (\ref{lam}) is not fine tuned 
neither to Kaluza-Klein scale $R^{-4}$ nor to the Planck scale $M_{pl(8)}^4$. 
Flat  Minkowsky geometry of the brane universe
is explained in our model by a special choice of matter equation of state,
rather than fine-tuning of the energy scale. 

Indeed, let us first consider the 
space-time ${\cal M}^8=R^4\times S^4$ without brane. 
If the geometry of $R^4$ sections of this space-time is Minkowsky,
${\cal M}^8$ is a solution 
of eight-dimensional Einstein equations with stress-energy tensor
(\ref{str}), (\ref{w}).
Such a stress-energy tensor can be generated, for example by a massless 
scalar field
$\Psi^a:{\cal M}^8\mapsto S^4$ which takes values on a four-dimensional
sphere. The stress-energy tensor of this field is
\begin{equation}
\label{psi}
T_{AB}=\Psi^a_{,A}\Psi^a_{,B}-\frac{1}{2}g_{AB}g^{CD}\Psi^a_{,C}\Psi^a_{,D}
\end{equation}
If we consider a topologically nontrivial configuration where 
$\Psi$ ``winds'' once on the 
sphere
\begin{equation}
\label{conf}
\Psi^a(x, y)=const\cdot y^a
\end{equation}
the parameter $w$ (\ref{w}) which describes the equation of state for 
$\Psi$ is   
\begin{equation}
\label{wpsi}
w_\Psi=1/2
\end{equation}  

Now, suppose that the geometry of the $R^4$ part of ${\cal M}^8$ is deSitter
or anti-deSitter with cosmological constant $\Lambda$, 
rather than Minkowsky. Such a space-time is a solution
of eight-dimensional Einstein equations with the stress-energy tensor of 
the form (\ref{str}) with energy density 
\begin{equation}
\rho=\frac{1}{8\pi G_8}\left(\Lambda+\frac{6}{R^2}\right)
\end{equation}
and equation of state
\begin{equation}
w=\frac{3+2\Lambda R^2}{6+\Lambda R^2}
\end{equation}
We see that if $\Lambda \not= 0$ 
the matter equation of state must be  different from (\ref{w}). 
Thus, if we take matter with equation of state  $w=1/2$, for example, 
the scalar field $\Psi$ in topologically nontrivial vacuum (\ref{conf}),
the four-dimensional sections of 
${\cal M}^8=R^4\times S^4$ will have flat geometry.

In the same way geometry of the brane is flat if the equation of 
state of matter in the region $r\le \hat r$ inside and on the boundary of 
the brane is adjusted in appropriate way.

 The size of space-time region 
around the brane in which the background metric is essentially deformed by 
the brane is very small $r_0\ll M_{pl(8)}^{-1}\ll R$  
even for a brane of very high tension.
This enables us to treat the brane as a test brane on a fixed 
background. 
The situation is analogous to the one encountered in particle physics where one
can neglect gravitational (self) interaction of particles because 
their gravitational radius is much smaller then the Compton wavelength. 
Fluctuations of the brane around its 
equilibrium position can be excited in particle collisions or astrophysical 
processes at the energy scales close to $\mu=\Lambda^{1/4}$ (\ref{lower})
 \cite{sundrum,dobado,creminelli,kugo}. 

We note also that cosmological models 
with a brane universe of nonzero tension must be very different from the 
conventional four-dimensional ones. Indeed, the large four-dimensional 
cosmological constant $\Lambda$ does not cause exponential expansion of the 
universe in our model. If we add some matter on the brane with equation 
of state $p=-\rho$ ($p$ and $\rho$ are four-dimensional pressure and energy 
density), the brane universe still can remain static, since we just change 
the brane cosmological constant $\Lambda\rightarrow\Lambda+\rho$ which would 
lead to the change of gravitational radius $r_0$ of the brane. The 
expansion of the brane universe is caused only by matter with equation of
state different from $p=-\rho$.      

I would like to thank V.Rubakov, I.Sachs and S.Solodukhin for many useful
discussions of the subject. The work was supported by SFB 375 
der DFG fur Astroteilchenphysik.

\end{document}